\documentclass[twocolumn,showpacs,amsmath,amssymb,pra,graphics,graphicx]{revtex4-1}
       \usepackage{graphicx}
        \usepackage{graphics}
\usepackage{bm}
\usepackage{dcolumn}
\begin{document}

\title{Moving  vortices in anisotropic superconductors}

\author{V. G. Kogan}
\email{kogan@ameslab.gov}
 \affiliation{Ames Laboratory--DOE, Ames, IA 50011, USA}
  \author{N. Nakagawa}
 \affiliation{Iowa State University, Ames, IA 50011, USA}   
 
  \date{\today}
       
\begin{abstract}
The magnetic   field  of moving vortices in anisotropic superconductors is considered  in the framework of time-dependent London approach. It is found that at distances large relative to the core size, the field may change sign that alludes to a non-trivial intervortex interaction which depends on the crystal anisotropy and on the speed and  direction of  motion. These effects are caused to the electric fields and corresponding normal currents which appear due to the moving vortex magnetic structure. We find that the   motion related part of the magnetic field attenuates at large distances as $1/r^3$ unlike the exponential decay of the static vortex field. The electric field induced by the vortex motion decreases as $1/r^2$. 
   \end{abstract}

\maketitle\section{Introduction}

The problem of interaction of vortices in anisotropic superconductors has been studied extensively in early 90s both theoretically \cite{Grishin,Buzdin,NakThiem} and experimentally \cite{Bolle}. For  vortices parallel to one of the principal crystal directions  the problem is solved just by  rescaling  the isotropic results. In particular the interaction is repulsive for any position of the second vortex relative to the first. However, the force direction in general is not along the vector $\bm R$ connecting the vortices, in other words, for an arbitrary positions of the pair there is a torque, unless $\bm R$ is directed along principal directions \cite{forces}. 

The situation is different if parallel vortices are tilted out of principal directions  \cite{Grishin,Buzdin,NakThiem}. Then, at distances of the order of London penetration depth $\lambda$, the  magnetic field $\bm h(\bm R)$  of a single tilted vortex
may change sign and approach zero for $R\to\infty$ being negative. In other words, the vortex-vortex interaction  being repulsive at short distances may turn attractive at large distances. This leads to formation of chains of vortices in tilted fields \cite{Bolle}.

In this paper we consider the magnetic field and current distributions of {\it moving} anisotropic vortices. Commonly, moving vortices are considered as static but displaced as a {\it whole}. It was argued, however, that out-of-core moving vortex structure differs from the static case due to out-of-core dissipation \cite{leo,TDL}. The moving vortex magnetic field $h(r,t)$ generates the electric field and currents of normal excitations, which in turn distort the field $h$. We show that at large distances the distortion is not small and even able to change the field sign. Unexpectedly, this distortion attenuates with distance as a power law $1/R^3$, i.e. much slower than the standard decay of undistorted field $\sim e^{-R/\lambda}$.  

At   distances large in comparison to the core size of interest in this work, one can use the time-dependent London approach based on the assumption that the current   consists of the normal and superconducting parts:
\begin{equation}
{\bm J}= \sigma {\bm E} -\frac{2e^2 |\Psi|^2}{mc}\, \left( {\bm
A}+\frac{\phi_0}{2\pi}{\bm
\nabla}\chi\right)  \,,\label{current}
\end{equation}
where  $\bm A$ is the vector potential, $\Psi$ is the order parameter,  $\chi$ is the phase,  $\phi_0$ is the flux quantum, ${\bm E}$ is the electric field, and $\sigma$ is the conductivity associated with normal excitations. 

The conductivity  $\sigma$ approaches the normal state value  $\sigma_n$
when the temperature $T$ approaches $T_c$; in  s-wave
superconductors  it vanishes   with decreasing temperature along with the density of normal excitations. This is not the case, however, for strong pair-breaking when   superconductivity is gapless while the density of states approaches the normal state value at all temperatures. Unfortunately,  not much experimental information about the $T$ dependence of $\sigma$ is available. Theoretically, this question is still debated, e.g.  Ref.\,\cite{Andreev} discusses possible enhancement of $\sigma$ due to inelastic scattering. Experimentally, interpretation of the microwave absorption data is not yet settled either \cite{Maeda}.

At distances large in comparison with the vortex core size, $|\Psi|$ is a constant $ \Psi_0 $ and Eq.\,(\ref{current}) becomes:
\begin{equation}
\frac{4\pi}{c}{\bm J}= \frac{4\pi\sigma}{c} {\bm E} -\frac{1}{\lambda^2}\,
\left( {\bm A}+\frac{\phi_0}{2\pi}{\bm
\nabla}\chi\right)  \,,
\label{current1}
\end{equation}
where $\lambda^2=mc^2/8\pi e^2|\Psi_0|^2 $ is the London penetration depth.
Acting on this by curl one obtains:
\begin{equation}
{\bm h}- \lambda^2\nabla^2{\bm h} +\tau\,\frac{\partial {\bm h}}{\partial
t}= \phi_0 {\bm z}\sum_{\nu}\delta({\bm r}-{\bm r_\nu})\,,\label{TDL}
\end{equation}
where   ${\bm r_\nu}(t) $ is the position of the $\nu$-th vortex which may depend on time $t$, $\bm z$ is the direction of vortices, and the relaxation time 
\begin{equation}
\tau=  4\pi\sigma\lambda^2/c^2 \,.
\label{tau}
\end{equation}
Equation (\ref{TDL}) can be considered as a general form of the time
dependent London equation (TDL). The anisotropic generalization of this equation was given in \cite{anisTDL} and reproduced here in Section III.
  
\section{Vortex at rest in anisotropic case}

For an arbitrary oriented vortex in anisotropic material this problem have been considered in \cite{K81,Grishin}. In general, results are cumbersome, so here we consider a simple situation of an orthorhombic superconductor in  field   along the $c$ axis.  The  London equation in this case is:
 \begin{eqnarray}
  h_z(x,y)- \lambda^2_1 \, \frac{\partial^2h_z}{\partial y^2}- \lambda^2_2  \, \frac{\partial^2h_z}{\partial x^2 } = \phi_0\delta(\bm r)\,,
 \label{hz} 
 \end{eqnarray} 
 Here, the frame $x,y,z$ is chosen to coincide with $a,b,c$ of the crystal, $\bm r=(x,y)$,   $\lambda^2_{xx}=\lambda^2_1$ and $\lambda^2_{yy}=\lambda^2_2$ are the diagonal components of the tensor $(\lambda^2)_{ik} $. The solution of this equation is     
\begin{eqnarray}
  h_z(x,y)=    \frac{ \phi_0 }{2\pi\lambda_1\lambda_ 2} K_0\left(\rho  \right)\,,\quad \rho^2=\frac{x^2}{\lambda_2^2} + \frac{y^2}{\lambda_1^2}\,.
 \label{h0} 
 \end{eqnarray}
Current densities follow:
\begin{eqnarray}
  J_x =   - \frac{c \phi_0 }{8\pi^2\lambda_1^3\lambda_2 } \frac{y\,K_1(\rho)}{\rho}\,,\quad  J_y =    \frac{ c\phi_0 }{8\pi^2\lambda_1\lambda_2^3 } \frac{x\,K_1(\rho)}{\rho}\,,\qquad
 \label{Jab} 
 \end{eqnarray}
where $K_{0,1}$ are Modified Bessel functions. 

\begin{figure}[h ]
\includegraphics[width=7cm] {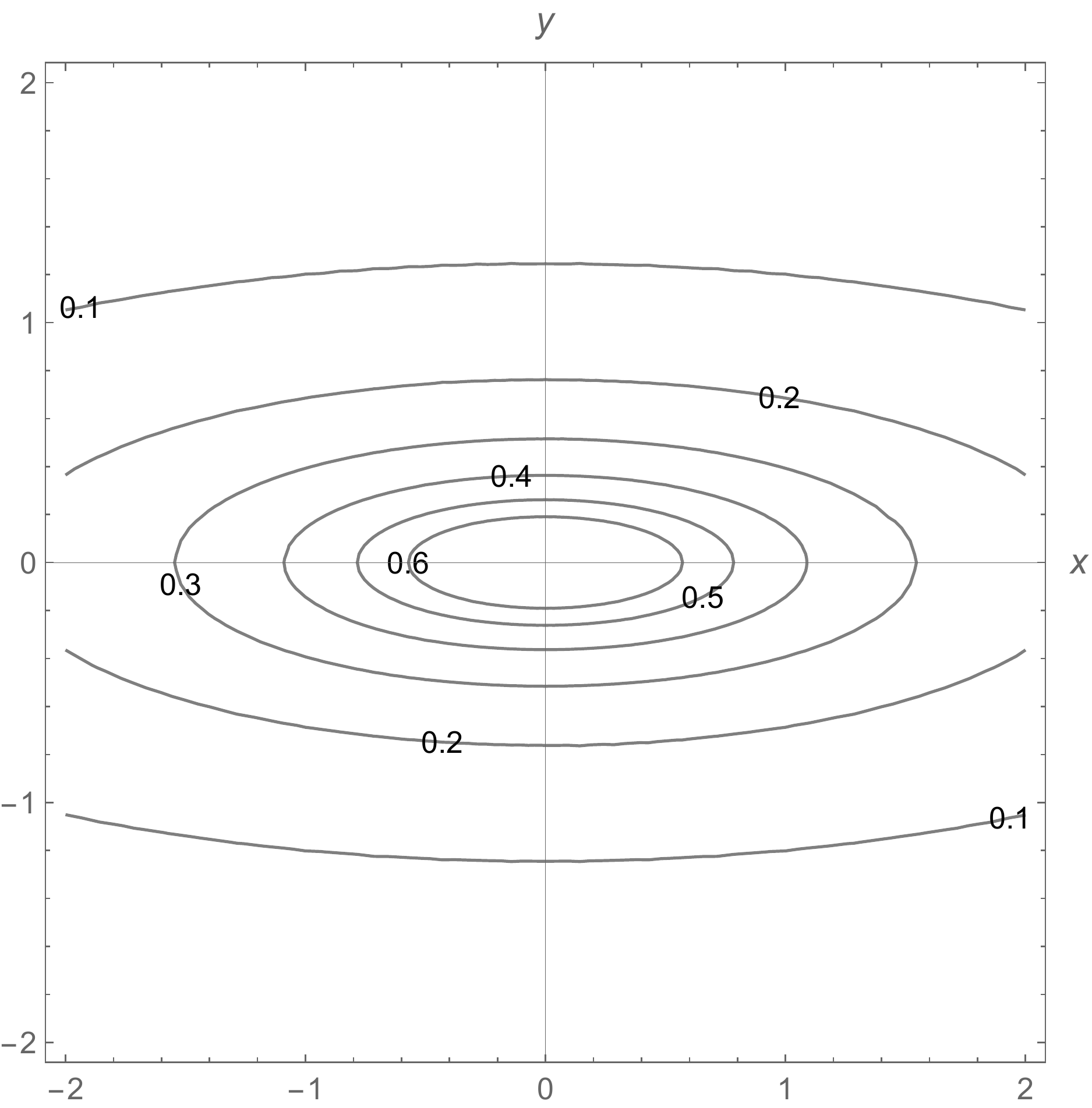}
\caption{The stream lines of the current for $\gamma=\lambda_2/\lambda_1=3$ or, which is the same, contours of constant $h_z(x,y)$.   $\lambda_1$ is taken as unit length. }
\label{f1a}
\end{figure}

It is easy to see that the contours $ h_z(x,y)= \,\,$const coincide with the stream lines of the current, an example is shown in Fig.\,\ref{f1a}.
The current lines have the expected ellipse-like shape. 
  \begin{figure}[h ]
\includegraphics[width=7cm] {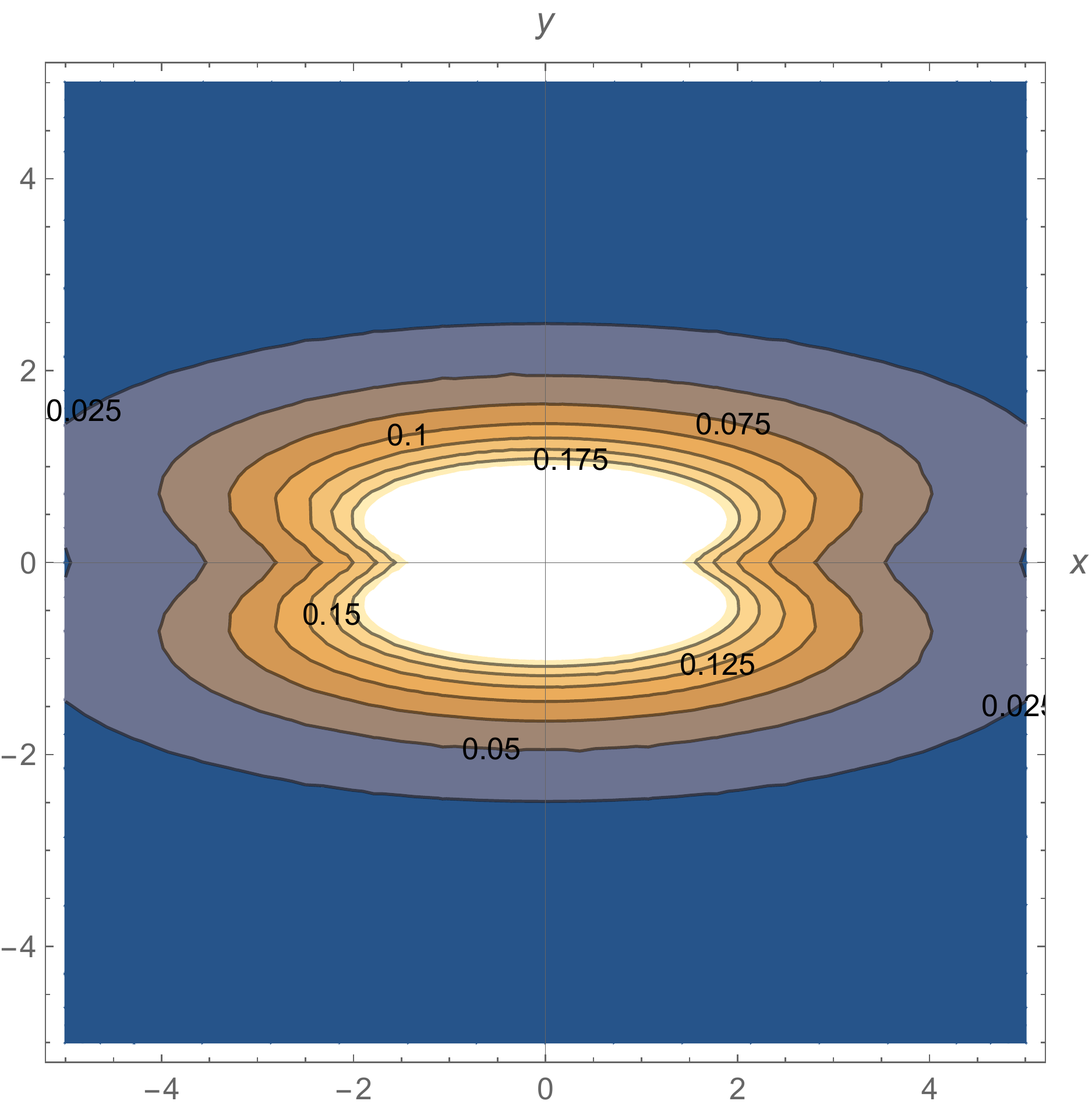}
\caption{The  contours of constant  current {\it values} $J(x,y)=\sqrt{J_x^2+J_y^2}$ for $\lambda_2/\lambda_1= 3$.    $x$ and $y$ are in units of $\lambda_1$. }
\label{f2a}
\end{figure}

This is, however, not the case for the distribution of the current {\it values} $J(x,y)=\sqrt{J_x^2+J_y^2}$.    An example is shown in Fig.\,\ref{f2a}. 
  Hence, the geometry of the streamlines of the vector $\bm J$ differs from that of contours $|J(x,y)|=const$, unlike the isotropic case where they are in fact the same. 
  
\section{Moving vortex}
 
The  anisotropic generalization of the isotropic Eq.\,(\ref{current1}) for the current is straightforward:
 \begin{equation}
 J_k= \sigma_{kl}  E_l -\frac{c}{4\pi}\left(\lambda^{-2}\right)_{kl}\left  (A_l + \frac{\phi_0}{2\pi} \frac{\partial\chi}{\partial x_l}\right)\,.
\label{current-a}
\end{equation}
Here, $\sigma_{kl}$ and $\left(\lambda^{-2}\right)_{kl}$ are tensors of the conductivity due to normal excitations and of the inverse square of the penetration depth.  

Having in mind to derive an equation for the magnetic field $\bm h$ we first have 
to get rid of the vector potential.  To this end, multiply  both sides by $4\pi \left(\lambda^{2}\right)_{k\mu}/c$ where $\left(\lambda^{2}\right)_{k\mu}$ is the tensor inverse to $\left(\lambda^{-2}\right)_{k\mu}$ and sum  up over $k$. Then  apply ${\rm curl}$ to both sides  and  use the relation  
\begin{equation}
{\rm curl} (\bm A +\phi_0 \bm \nabla \chi/2\pi)= \bm h-\phi_0\hat{\bm z}\delta(\bm r-\bm r_\nu)\,,
\end{equation}
where $\bm r_\nu$ is the vortex core position. 
 
 It is convenient to use in the following the notation curl$_\nu\bm V= \epsilon_{\nu s\mu}\partial V_\mu/\partial x_s$ where $\epsilon_{\nu s\mu }$ is Levi-Chivita unit antisymmetric tensor:  $\epsilon_{xyz}=1$ and so do all components  with even number of transpositions of indices,   it is $-1$ for  odd numbers, and zero otherwise. 
 
 Hence, applying $\epsilon_{\nu s\mu}\partial /\partial x_s$ to Eq.\,(\ref{current-a}), one obtains the anisotropic version of TDL \cite{anisTDL}:
 \begin{eqnarray}
h_\nu &+&\frac{4\pi}{c}\epsilon_{\nu s\mu} \lambda^2_{k\mu} \frac{\partial J_k}{\partial x_s} - \frac{4\pi}{c}\epsilon_{\nu s\mu} \lambda^2_{k\mu}\sigma_{kl} \frac{\partial E_l}{\partial x_s} \nonumber\\
&=&\phi_0 \hat{\bm z}_\nu\delta(\bm r-\bm v t).\qquad
\label{anis-TDL}
\end{eqnarray}
In this form, the equation is valid for an arbitrary oriented vortex and  any crystal anisotropy. 

For an orthorhombic crystal in which the vortex and its field are along one of the principal directions (call it $z$), this cumbersome equation takes the form: 
 \begin{eqnarray}
h_z &-&\frac{4\pi}{c}\left( \lambda^2_{xx} \frac{\partial J_x}{\partial y} -  \lambda^2_{yy} \frac{\partial J_y}{\partial x}\right)
\nonumber\\
&+&\frac{4\pi\sigma}{c}\left( \lambda^2_{xx} \frac{\partial E_x}{\partial y}  -  \lambda^2_{yy}  \frac{\partial E_y}{\partial x}\right)=\phi_0  \delta(\bm r-\bm v t).\qquad
\label{ortho-TDL}
\end{eqnarray}
Here we further simplified the problem assuming isotropic conductivity of normal excitations $\sigma_{xx}=\sigma_{yy}=\sigma $.
This should be solved together with quasi-stationary Maxwell equations curl$\bm E =-\partial_t\bm h/c$ and div$\bm E =0$ \cite{LL,Gorkov}, which can be done  in 2D Fourier space:
 \begin{equation}
   E_{\bm k x}=-\frac{k_y}{k_x}E_{\bm k y}= -\frac{ik_y}{ck^2} \,\frac{\partial h_{\bm k z}}{\partial t}  \,,
\label{Es}
\end{equation}
so that we obtain the 2D Fourier transform of Eq.\,(\ref{ortho-TDL}):
 \begin{eqnarray}
h_{\bm k}   &&  \left(1+k_x^2 \lambda^2_{yy}+ k_y^2 \lambda^2_{xx}\right) \nonumber\\
&&
+\frac{4\pi\sigma}{c^2}\,\frac{ \lambda^2_{yy}k_x^2+ \lambda^2_{xx}k_y^2}{k^2}
  \frac{\partial h_{\bm k}   }{\partial t} =\phi_0e^{-i\bm k\bm v t}\,,
 \label{FTortho-TDL}
\end{eqnarray}
where $h_{\bm k}  $ is the Fourier transform of $h_z (\bm r)$. In isotropic case we obtain the equation studied in \cite{TDL}. We further denote 
$\lambda^2_{yy}=\lambda_2^2,\quad \lambda^2_{xx}=\lambda_1^2$  and $\lambda=\sqrt{\lambda_1\lambda_2}$.   The anisotropy parameter is defined as $\gamma=\lambda_2/\lambda_1$. Then, we obtain:
 \begin{eqnarray}
h_{\bm k}   \left(1+k_x^2 \lambda^2_2+ k_y^2 \lambda^2_1\right) 
+\tau \,\frac{ \lambda^2_2k_x^2+ \lambda^2_1k_y^2}{\lambda^2k^2}
  \frac{\partial h_{\bm k}  }{\partial t} =\phi_0e^{-i\bm k\bm v t}. \nonumber\\
  \label{ h(k,t)}
\end{eqnarray}
with $\tau=4\pi\sigma\lambda^2/c^2$. 
This is a linear differential equation for $h_{\bm k}(t)$ with the solution
\begin{eqnarray}
&&h_{\bm k} =\frac{\phi_0e^{-i\bm k\bm v t}}{ C-iD\bm k\cdot\bm s}\,,\quad \bm s=  \bm v\tau \,,
\nonumber\\
&& C=1+k_x^2 \lambda^2_2+ k_y^2 \lambda^2_1\,,\quad D=\frac{ \lambda^2_2k_x^2+ \lambda^2_1k_y^2}{\lambda^2k^2}\,. \qquad
 \label{h(k,t)}
\end{eqnarray}
 Since we are interested in stationary motion with a constant velocity, we can set here $t=0$.  
 
 The dimensionless parameter  
\begin{eqnarray}
S=\frac{s}{\lambda}=\frac{4\pi v\sigma\lambda}{c^2} 
  \label{ S}
\end{eqnarray}
is small even for vortex velocities exceeding the speed of sound presently attainable  \cite{Eli,Denis}. 
  Although in principle $S$ can take larger values, we restrict this discussion by small $S$ and  call this case a ``slow motion".

\section{Slow motion}

For $s\to 0$  one can expand $h(\bm k, \bm s)$ in powers of small $s$ up to ${\cal O}(s)$:
\begin{eqnarray}
 h_{\bm k}=\frac{\phi_0 }{ C}+i\frac{\phi_0D}{C^2}\bm k\cdot\bm s\,, 
  \label{ expand}
\end{eqnarray}
The first term corresponds to the static solution discussed above:
 \begin{eqnarray}
  h_0(x,y)=    \frac{ \phi_0 }{2\pi\lambda^2} K_0\left(\rho  \right)\,,\quad \rho^2=\frac{x^2}{\lambda_2^2} + \frac{y^2}{\lambda_1^2}\,.
 \label{h0} 
 \end{eqnarray}
 The correction due to motion is given by
 \begin{eqnarray}
 \frac{\delta h_{\bm k}\lambda^2}{\phi_0}= i\frac{(\lambda^2_2k_x^2+ \lambda^2_1k_y^2)\bm k\cdot\bm s}{k^2(1+\lambda^2_2k_x^2+ \lambda^2_1k_y^2)^2}\,, 
  \label{corr}
\end{eqnarray}
To separate the part that does not disappear when  $\lambda_1= \lambda_2$, one can use the identity
\begin{eqnarray}
   \frac{ \lambda^2_2k_x^2+ \lambda^2_1k_y^2 }{ k_x^2+k_y^2 }=\lambda_2^2+
   \frac{ k_y^2(\lambda^2_1 - \lambda^2_2) }{ k_x^2+k_y^2 }  
  \label{ident}
\end{eqnarray}
to obtain:
 \begin{eqnarray}
 &&\frac{4\pi^2\lambda^2\delta h(\bm r)}{i\phi_0}= \lambda_2^2\int  \frac{d^2\bm k(\bm k\cdot\bm s)e^{i\bm k \bm r}} 
 {(1+\lambda^2_2k_x^2+ \lambda^2_1k_y^2)^2}\nonumber\\
 &&+(\lambda^2_1 - \lambda^2_2)\int  \frac{d^2\bm kk_y^2(\bm k\cdot\bm s) e^{i\bm k \bm r}} 
 {k^2(1+\lambda^2_2k_x^2+ \lambda^2_1k_y^2)^2}  \,. 
  \label{corr1}
\end{eqnarray}

Evaluation of the first contribution is outlined in Appendix A:  
\begin{eqnarray}
h_1 = -  \frac{\phi_0}{2\pi\lambda^2} \frac{  S_xX +S_yY\gamma^2}{2}   K_0\left(\sqrt{\frac{X^2}{\gamma}+ Y^2\gamma} \right)\,, \qquad
  \label{h1}
\end{eqnarray}
 where
\begin{eqnarray}
   \bm S =\frac{\bm s }{\lambda},\,\,\,  X=\frac{x}{\lambda},\,\,\, Y=\frac{y}{\lambda},\,\,\, \lambda=\sqrt{\lambda_1\lambda_2},\,\,\, \gamma=\frac{\lambda_2}{\lambda_1}.\qquad
  \label{eq23}
\end{eqnarray}

It is shown in \cite{Norio2} that in the isotropic case for a vortex moving along $x$
 \begin{eqnarray}
  h (\bm r)   
 = \frac{\phi_0}{2\pi\lambda^2} e^{-sx/2\lambda^2} K_0\left(\frac{r}{2\lambda}\sqrt{4+s^2/\lambda^2}\right) \qquad
\label{hz(r)1b} 
\end{eqnarray}
in common units. 
Expanding this in small $s$ one obtains for a slow motion:
 \begin{eqnarray}
 \delta h (\bm r)   
 =- \frac{\phi_0}{4\pi\lambda^4}  s x  K_0\left(\frac{r}{\lambda}\right)\,.
\label{hz(r)1c} 
\end{eqnarray}
Hence,   $h_1$ of Eq.\,(\ref{h1}) has the correct isotropic limit.

The    second  integral over two components of $\bm k$ in Eq.\,(\ref{corr1})    can be reduced to  integrals over a single variable which are easy to deal with numerically, see Appendix B:
\begin{widetext}
 \begin{eqnarray}
&&\frac{2\pi\lambda^2}{\phi_0} h_2= \frac{ ( \gamma^2-1)}{4\gamma}\Big\{S_xX\int_0^\infty \frac{d\zeta}{(\zeta+\gamma)^{3/2}(\zeta+1/\gamma)^{3/2}} 
 \left[  
 K_0\left( {\cal R}_\zeta\right)-\frac{Y^2}{(\zeta+1/\gamma) {\cal R}_\zeta} K_1\left( {\cal R}_\zeta\right)    \right] \nonumber\\
 &&+S_yY\int_0^\infty \frac{d\zeta}{(\zeta+\gamma)^{1/2}(\zeta+1/\gamma)^{5/2}} 
 \left[  
 3K_0\left( {\cal R}_\zeta\right)-\frac{Y^2}{(\zeta+1/\gamma) {\cal R}_\zeta} K_1\left( {\cal R}_\zeta\right)    \right]\Big\},   \qquad
 {\cal R}_\zeta=\sqrt{\frac{X^2}{\zeta+\gamma}+\frac{Y^2}{\zeta+1/\gamma}}\,.\qquad\qquad\qquad\qquad
  \label{dh2}
\end{eqnarray}

 \end{widetext}
 
Thus, the vortex field can be calculated as $h=h_0+h_1+h_2$ with $h_0$ given in Eq.\,(\ref{h0}), $h_1$ in Eq.\,(\ref{h1}),  and $h_2$ in Eq.\,(\ref{dh2}). The results obtained with the help of Wolfram Mathematica package are shown below.
  
  \begin{figure}[h ]
\includegraphics[width=7cm] {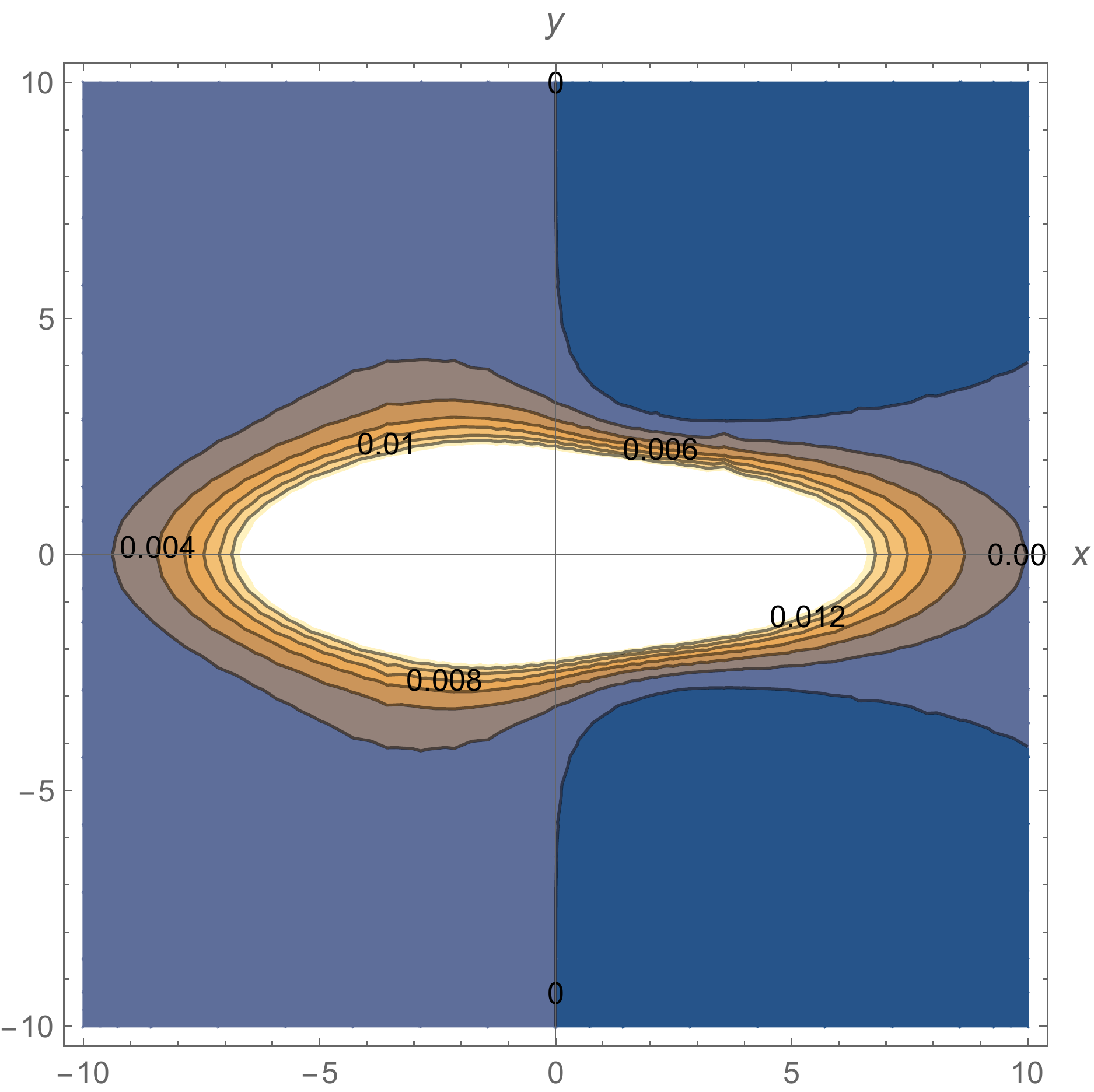}
\caption{Contours $h(x,y)=$ const   for the vortex moving along $x$ axis ($S_x=0.1, \,\,S_y=0$) and $\lambda_2/\lambda_1= 3$.  The motion is directed to $+x$.  $x$ and $y$ are in units of $\lambda=\sqrt{\lambda_1\lambda_2} $. }
\label{f3}
\end{figure}

One can see in Fig.\,\ref{f3} that the current stream-lines (or, what is the same, contours $h(x,y)=$ const) in the vicinity of the moving vortex core are only weakly distorted relative to the static elliptic shape.  The most interesting feature of this distribution is that at large distances $h(x,y)$ changes sign in some parts of the $(x,y)$ plane. Since the interaction energy of the vortex at the origin with another one at $(x,y)$ is proportional to $h(x,y)$, the presence of domains with $h<0$ means that for the second  vortex  in these domains the intervortex interaction is attractive. 
 
\begin{figure}[h ]
\includegraphics[width=7cm] {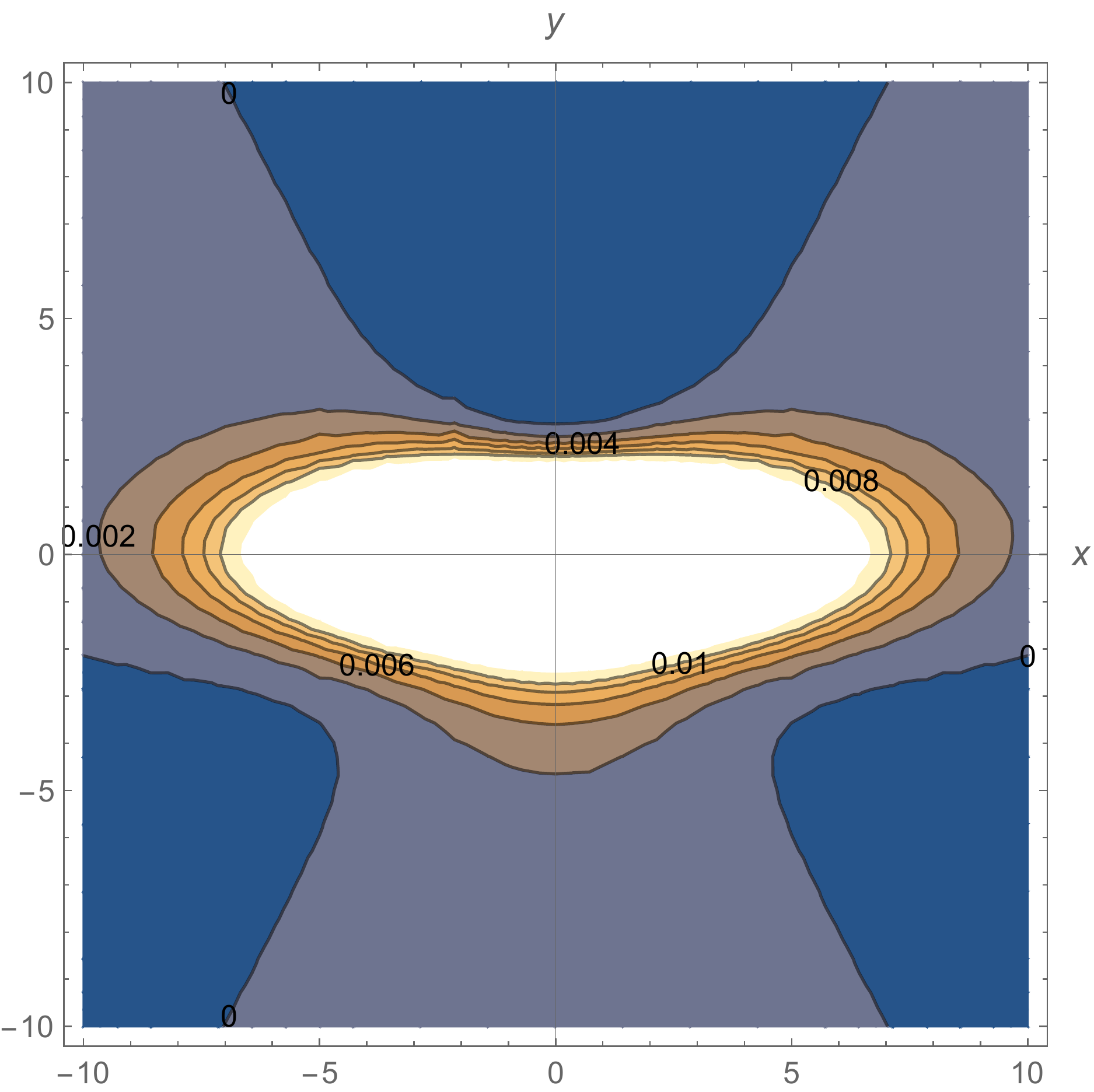}
\caption{Contours $h(x,y)=$ const   for the vortex moving along $y$ axis ($S_x=0, S_y=0.1$) and $\lambda_2/\lambda_1= 3$. The motion is directed to $+y$.   $x$ and $y$ are in units of $\lambda=\sqrt{\lambda_1\lambda_2} $.  }
\label{f4}
\end{figure}

The field distribution is different for the motion along $y$ axis shown in Fig.\,\ref{f4}.
  \begin{figure}[htb ]
\includegraphics[width=7cm] {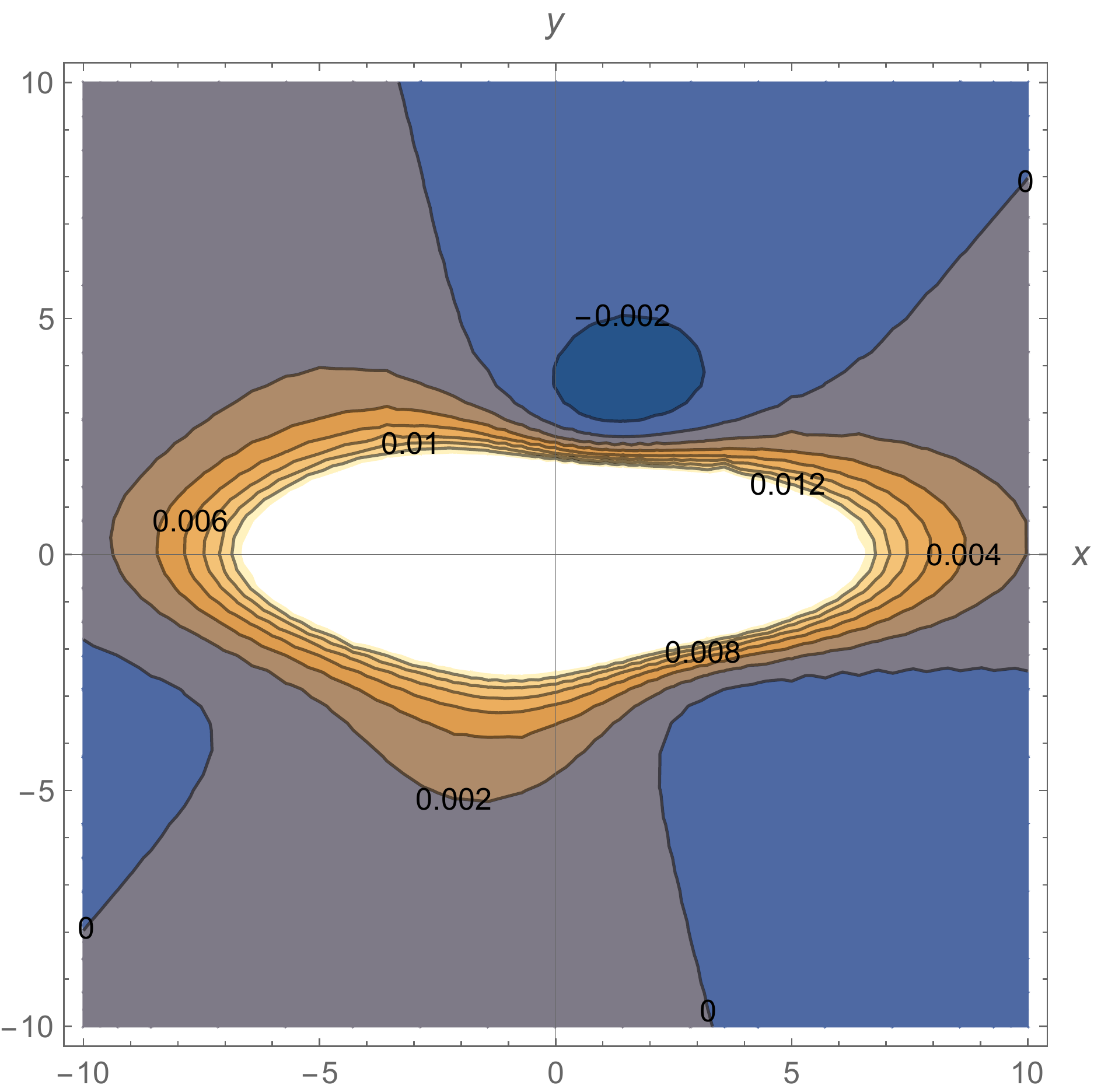}
\caption{Contours $h(x,y)=$ const   for the vortex moving along the diagonal $x=y$ ($S_x=S_y=0.1$) and $\lambda_2/\lambda_1= 3$.    $x$ and $y$ are in units of $\lambda=\sqrt{\lambda_1\lambda_2} $. }
\label{f5}
\end{figure}

It is  seen that the flux in front of the moving vortex is depleted whereas behind it is enhanced, the feature first discussed in \cite{norio1} for the isotropic case. 
This feature remains also for a general  direction of motion; an example of motion along the line $x=y $ is shown in  Fig.\,\ref{f5}. Moreover, Fig.\,\ref{f3}--\ref{f5} show  that this depletion may even change sign of the field. 

It is worth mentioning that the London theory is  reliable in the region $r\gg \xi$, $\xi$ being the core size, and so are  our predictions of a non-trivial behavior of $h(x,y)$ at large distances. 
  \begin{figure}[h]
\includegraphics[width=7cm] {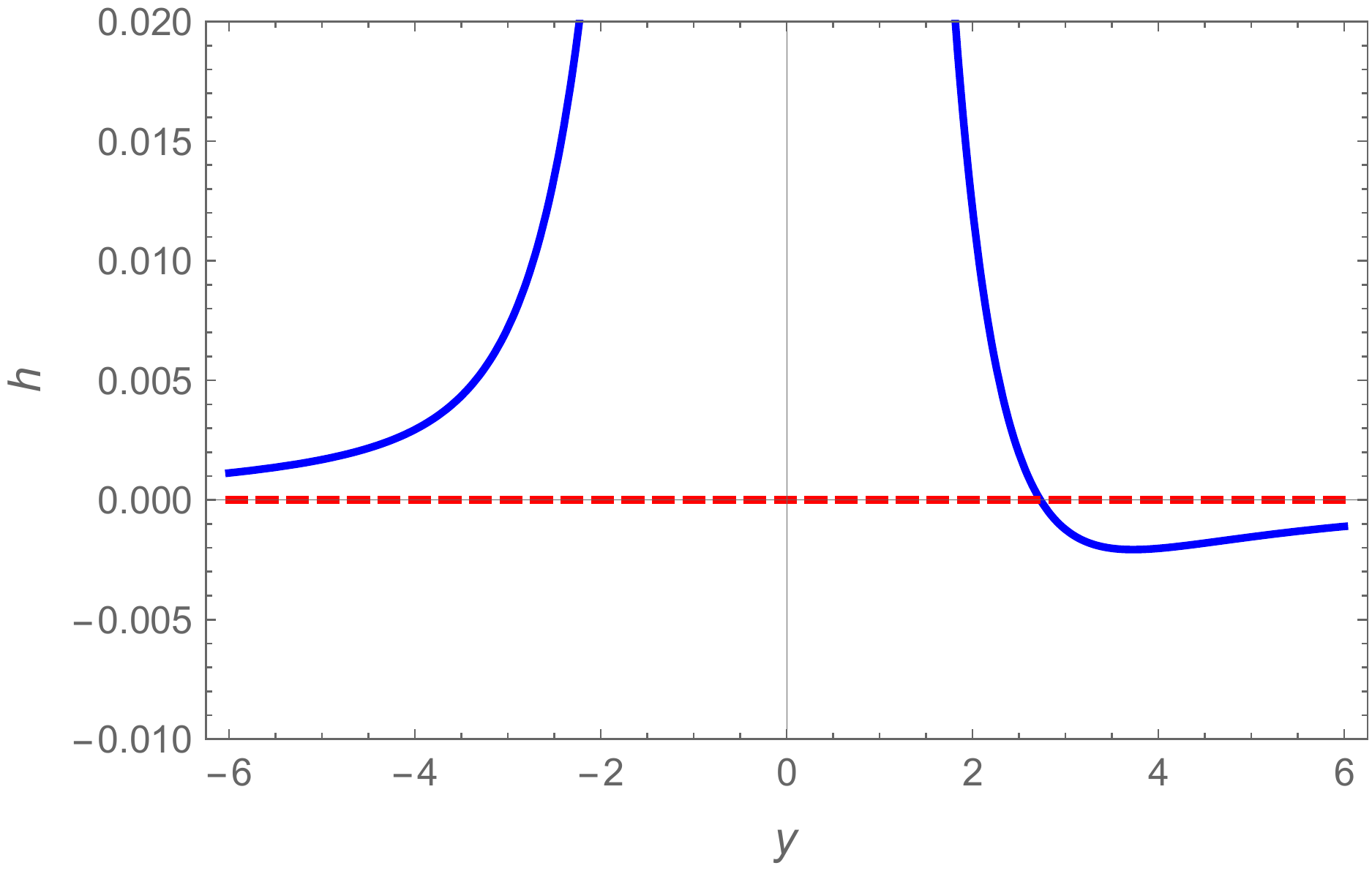}
\caption{The field $h(0,y)$  for the vortex moving along  $y$ ($S_x=0, S_y=0.1 $); $\lambda_2/\lambda_1= 3$.     $x$ is in units of $\lambda=\sqrt{\lambda_1\lambda_2} $. }
\label{f6}
\end{figure}

It is instructive to see how the interaction changes along certain directions. E.g., for $S_x=0, \,\,\, S_y=0.1$,  the motion along $y$-axis, $h(0,Y)$ is positive if $0<Y\lesssim 2.5$ (so that the second vortex at $(0,Y)$ in this region is repelled by the vortex at the origin). If the second vortex is at $2.5\lesssim Y<\infty$ the interaction is attractive. This is illustrated  in Fig.\,\ref{f6}.

  \begin{figure}[htb ]
\includegraphics[width=7cm] {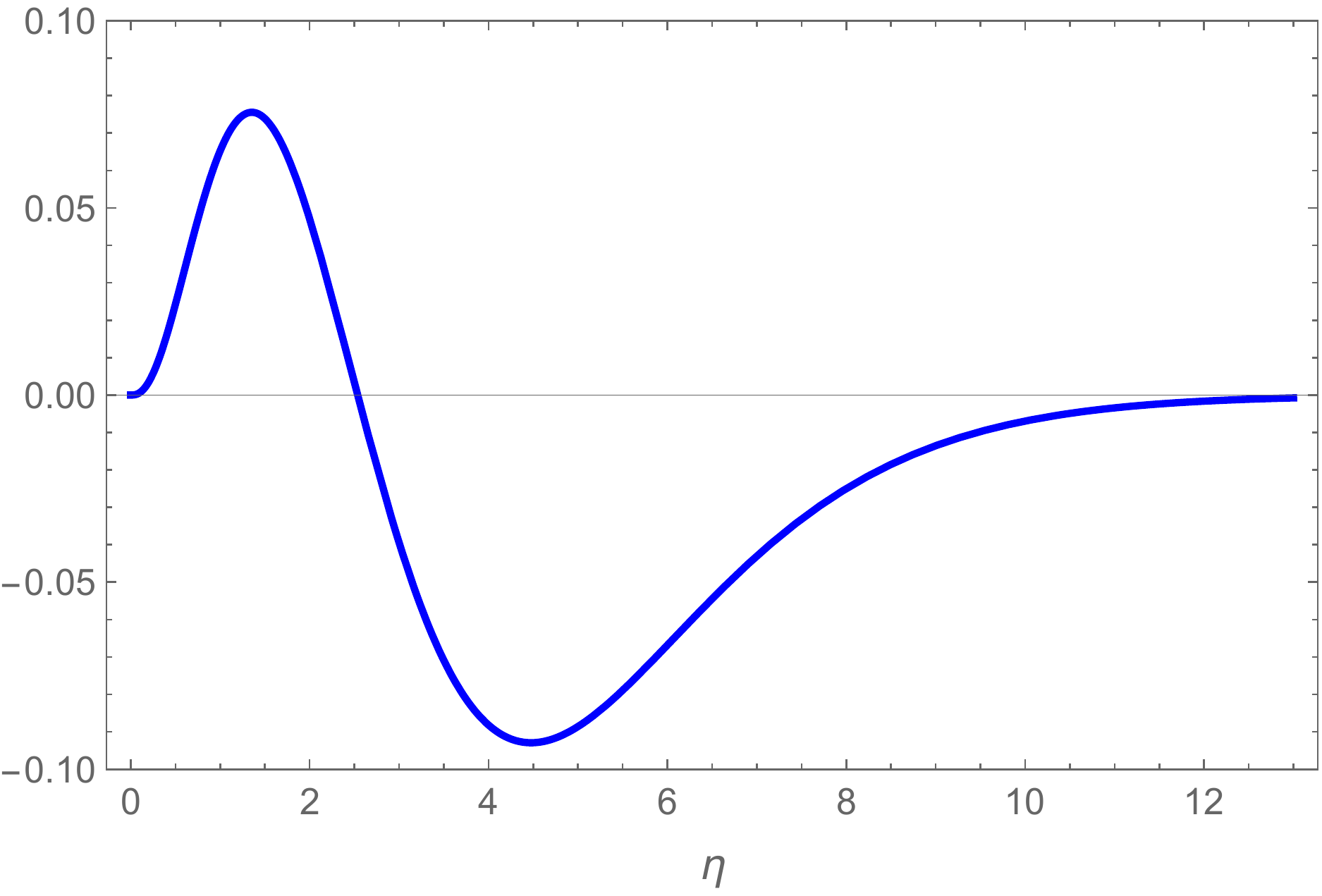}
\caption{The integrand of Eq.\,(\ref{int R}) for $Y=10$ and $\gamma=3$.  }
\label{f7}
\end{figure}

\subsection{Asymptotic behavior of $\bm {h (0,Y) }$ for $\bm{Y\to\infty}$}

For $X=0$, Eq.\,(\ref{dh2}) yields
 \begin{eqnarray}
 \frac{2\pi\lambda^2}{\phi_0} h_2=   \frac{ \gamma^2-1}{4\gamma} S_y Y \int_0^\infty \frac{d\zeta\left[3K_0( \eta )-  \eta K_1 (\eta)\right]}{(\zeta+\gamma)^{1/2}(\zeta+1/\gamma)^{5/2}}, 
 \nonumber\\
\eta=\frac{|Y|}{ \sqrt{\zeta+1/\gamma}}\,.\qquad\qquad\qquad\qquad
  \label{x=0}
\end{eqnarray}
 Going to the integration variable $\eta$, we get  
 \begin{eqnarray}
 \frac{2\pi\lambda^2}{\phi_0} h_2=    \frac{ \gamma^2-1}{2\gamma}
 \frac{S_y}{Y^2} \int_0^{Y\sqrt{\gamma}} \frac{d\eta\, \eta^3\left[3K_0(\eta)- \eta K_1 (\eta)\right]}{\sqrt{Y^2+\eta^2(\gamma-1/\gamma) }}.\nonumber\\
   \label{int R}
\end{eqnarray}
Fig.\,\ref{f7} shows that the    integrand here  is substantial only in the finite region $ 0< \eta\lesssim 10$. Therefore being interested in the asymptotic behavior   for $|Y|\to\infty$, one can replace the denominator by $|Y |$ and the upper limit of integration by $\infty$:
 \begin{eqnarray}
 \frac{2\pi\lambda^2}{\phi_0} h_2(0,Y)&=&    \frac{ \gamma^2-1}{2\gamma}
 \frac{ S_y}{Y^3} \int_0^{\infty}  d\eta\, \eta^3\left[3K_0 -  \eta K_1 \right]_\eta \qquad \nonumber\\
& =&  -  \frac{\gamma^2-1}{ \gamma}  
 \frac{2S_y}{Y^3} \,.
    \label{ass}
\end{eqnarray}
Thus, $h_2(0,Y)$ is negative when $Y\to\infty$ and positive for $Y\to -\infty$. It decays as $1/Y^3$, therefore,  the total field $h_0+h_1+h_2$ attenuates as a power law as well, since $h_0$ and $h_1$ decay exponentially and at large distances can be disregarded. Hence, $h_2$ can be replaced with $h$ in this region. 
This conclusion agrees with direct numerical evaluation of $h(0,Y)$ shown in Fig.\,\ref{f6}

In the same way one can obtain the leading term in the asymptotic behavior for $Y=S_y=0$ for the motion along the $x$ axis:
 \begin{eqnarray}
h(X,0)\sim   \frac{\phi_0}{2\pi\lambda^2}  \frac{ \gamma^2-1}{2\gamma}
 \frac{ 2S_x}{X^3} \,. 
    \label{assX}
\end{eqnarray}
For the sake of brevity we do not provide other terms in the asymptotic series.

The power-law decay of the field $h(x,y)$ for vortices moving in anisotropic superconductors is a surprising feature. Clearly, this feature disappears for vortices at rest as well as for vortices moving in isotropic materials.
Formally, the power-law behavior in real space  
originates in the factor $1/k^2$ in Fourier transforms, see e.g.  Eq.\,(\ref{corr1}), which, however, cancels out for $\gamma=1$.

\section{Electric field for slow motion}

In the approximation  linear in  velocity, we have according to Eq.\,(\ref{h(k,t)})
  \begin{eqnarray}
 \frac{\partial h_{\bm k}}{\partial t}   =-i \frac{\phi_0(\bm k\cdot\bm v)}{C}\,,\quad C=1+k_x^2 \lambda^2_2+ k_y^2 \lambda^2_1\,.
   \label{dh/dt}
\end{eqnarray}
This yields the electric field
 \begin{equation}
   E_{\bm k x}=-\frac{k_y}{k_x}E_{\bm k y}= -\frac{\phi_0}{c\tau}\frac{k_y(\bm k\cdot\bm s)}{ k^2C} \,,\quad \bm s =\bm v\tau\,,
   \label{Es1}
\end{equation}
see Eqs.\,(\ref{Es}). 
Hence, we have in real space
 \begin{equation}
   E_x =  -\frac{\phi_0}{4\pi^2c\tau }\int \frac{d^2\bm k\,k_y (\bm k\cdot\bm s)}{k^2C} e^{i\bm k \bm r}\,,    \label{Ex}
\end{equation}
or, using $\lambda=\sqrt{\lambda_1\lambda_2}$ as the unit length, 
 \begin{equation}
   E_x =  -\frac{\phi_0}{4\pi^2c\tau\lambda }\int \frac{d^2\bm q\,q_y (\bm q\cdot\bm S)}{ q^2C} e^{i\bm q \bm R}\,.   
    \label{Ex1}
\end{equation}
Here, $\bm q=\bm k\lambda$, $\bm R=(X,Y)=\bm r/\lambda$ (see definitions (\ref{eq23}), and 
\begin{equation}
   C = 1+q_x^2 \gamma+ q_y^2 /\gamma\,,\quad \gamma=\lambda_2/\lambda_1\,. 
    \label{C1}
\end{equation}
In the same way we obtain
 \begin{equation}
   E_y =   \frac{\phi_0}{4\pi^2c\tau\lambda }\int \frac{d^2\bm q\,q_x (\bm q\cdot\bm S)}{ q^2C} e^{i\bm q \bm R}\,.   
    \label{Ey1}
\end{equation}
The integrals in Eqs.\,(\ref{Ex1})  and (\ref{Ey1}) are dimensionless. 

It is of interest to see the streamlines of $\bm E$ (or, that is the same, of the normal current $\bm J_n=\sigma\bm E$). To this end, we calculate the stream function $G(x,y)$ such that $E_x=\partial_yG$ and 
$E_y=-\partial_xG$; the streamlines then are given by contours $G(x,y)= const$. In Fourier space we have $E_{x\bm k}=ik_y G_{\bm k}$ so that
 \begin{eqnarray}
   G_{\bm k}=  \frac{i\phi_0}{c\tau}\frac{ (\bm k\cdot\bm s)}{ k^2C},\,\,\,\, 
       G(\bm r)=  \frac{i\phi_0}{4\pi^2c\tau}\int  \frac{d^2\bm q   (\bm q\cdot\bm S)e^{i\bm q \bm R}}{ q^2C} .\qquad 
   \label{Gr}
\end{eqnarray}
 
The formal procedure of reducing the double to single integration in Eq.\,(\ref{Gr}) is similar to that used for $h(\bm r)$ and  is outlined  in Appendix C. The result is:
\begin{eqnarray}
       &&G(\bm r) = - \frac{\phi_0}{4\pi c\tau} \int_0^\infty    \frac{d\eta \, K_0({\cal R} \sqrt{\eta})  }{\sqrt{\mu\nu}}    \left( \frac{ S_xX }{ \mu}+\frac{S_y Y }{ \nu} \right),\nonumber\\ 
&& \mu= 1+\eta \gamma\,,\quad \nu= 1+\eta /\gamma\,,\quad {\cal R}=\sqrt{ 
    \frac{ X^2}{\mu}+\frac{ Y^2}{ \nu} }\,. \qquad 
                  \label{Gr1}
\end{eqnarray}
 
\begin{figure}[h ]
\includegraphics[width=7cm] {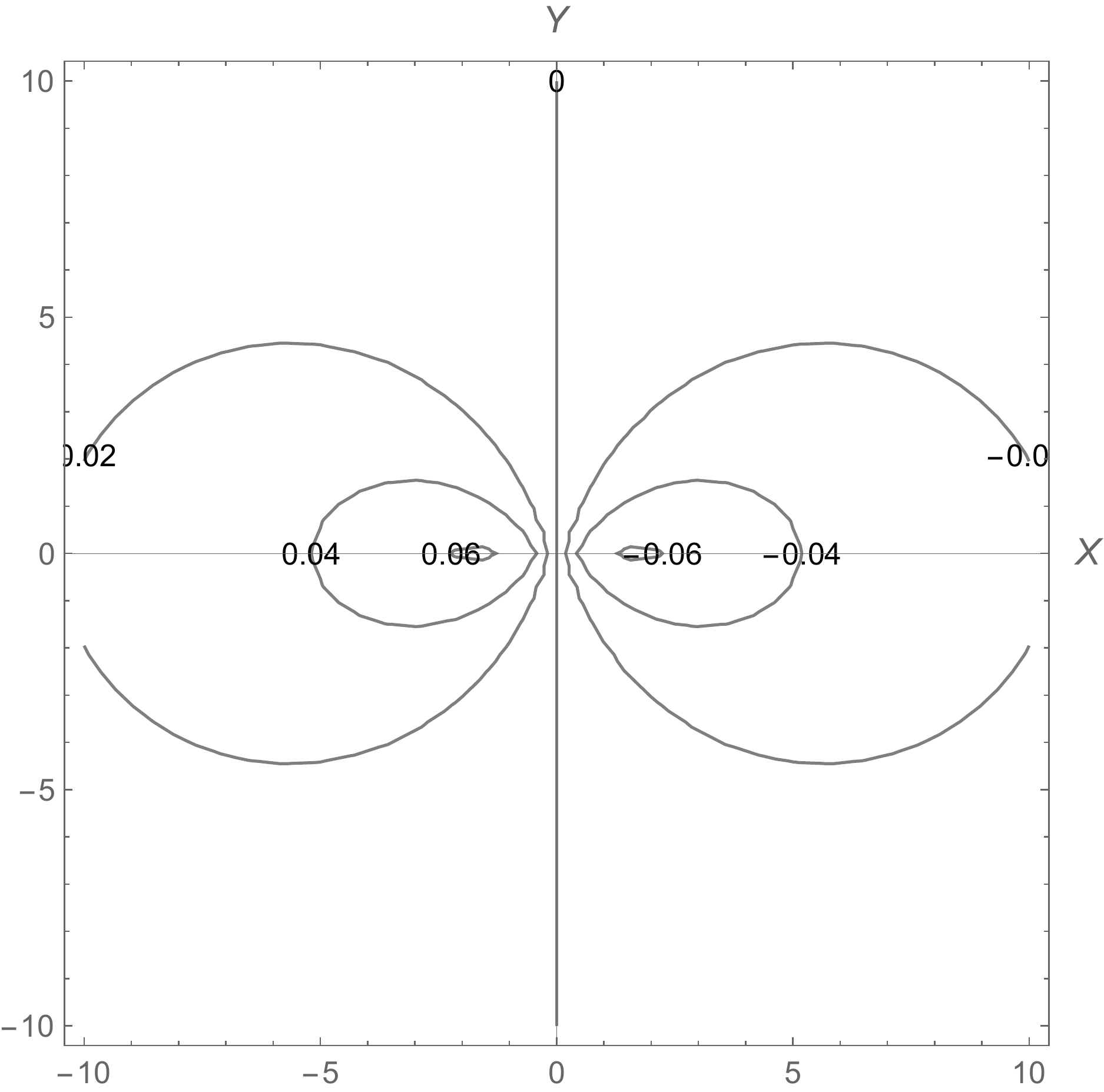}
\caption{ Streamlines of the field $\bm E$ (or of the normal current $\bm J_n$) for the vortex moving along  $X$ ($S_x=0.1, S_y=0$). $\gamma=\lambda_2/\lambda_1= 3$.    $X,Y$ are in units of $\lambda=\sqrt{\lambda_1\lambda_2} $. Positive  constants by the contours correspond to the clockwise current direction, negative otherwise.}
\label{f8}
\end{figure}

\begin{figure}[htb]
\includegraphics[width=7cm] {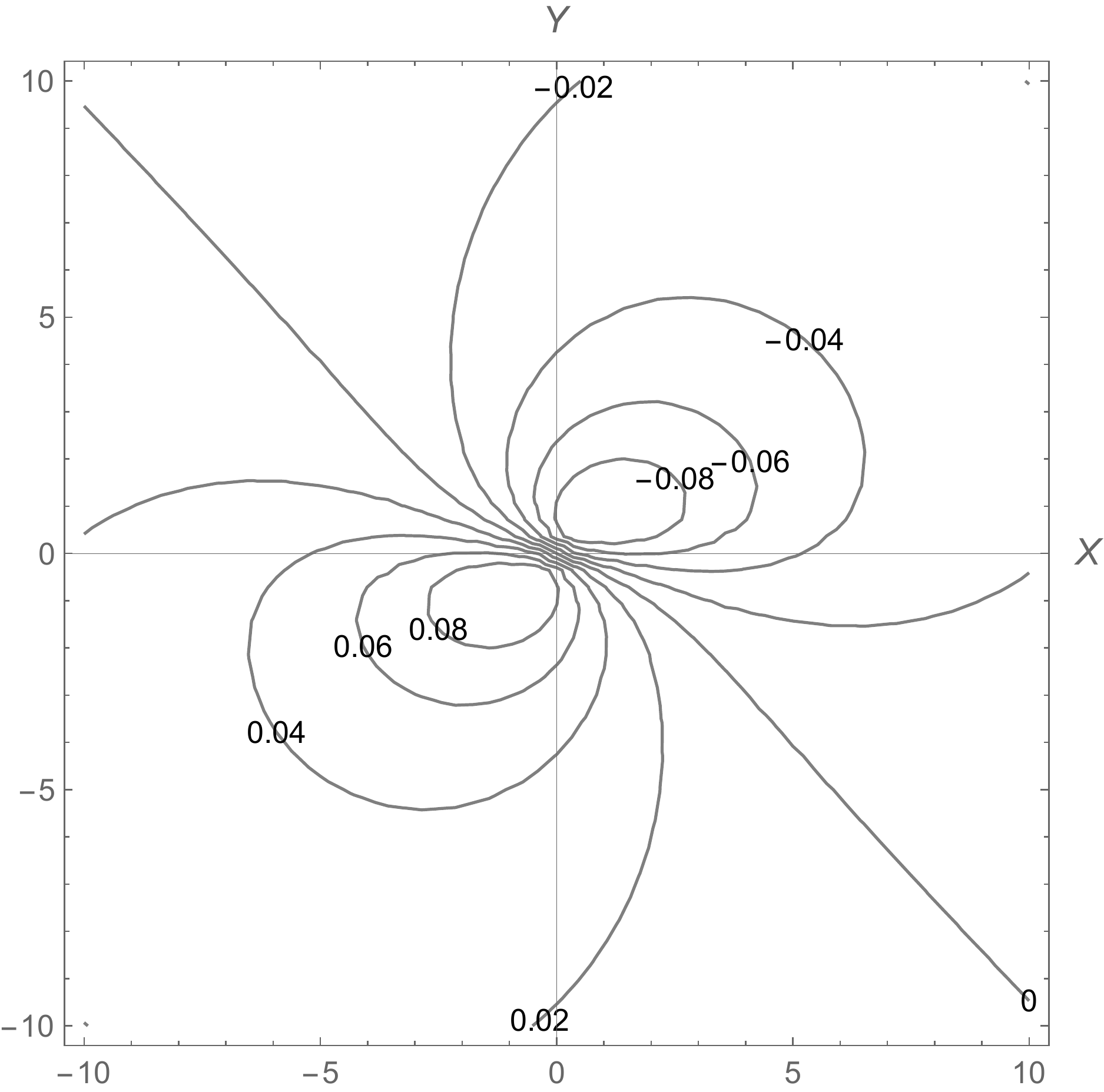}
\caption{ Streamlines of the normal current   for the vortex moving along    the line $X=Y$ ($S_x= S_y=0.1$), $\gamma=\lambda_2/\lambda_1= 3$.   $X,Y$ are in units of $\lambda=\sqrt{\lambda_1\lambda_2} $. Positive  constants by the contours correspond to the clockwise current direction, negative otherwise.}
\label{f9}
\end{figure}

 Figs.\,\ref{f8} and \ref{f9}  show  two examples of $J_n$-streamlines (or contours $G(X,Y)=\,\,$const) obtained by numerical integration of Eq.\,(\ref{Gr1}).  
 
The electric field  is now readily obtained by differentiation of $G$. We will not write down these cumbersome expressions. Instead we consider the asymptotic behavior of electric fields at large distances   in two relatively simple cases using the  method  employed above for asymptotic behavior of $h(0,y)$ and $h(x,0)$. Omitting formalities, we give the results:
\begin{eqnarray}
  G(X,0) \sim - \frac{\phi_0 }{2\pi c\tau }  \frac{S_x} {X}\,,\qquad |X|\to\infty\,,           
        \label{G(X)as)}
\end{eqnarray}
that yields
\begin{eqnarray}
  E_x(X,0)=0\,,\quad E_y(X,0) \sim  \frac{\phi_0 }{2\pi c\tau }  \frac{S_x} {X^2} \,.   
          \label{G(X)as)}
\end{eqnarray}
Similarly,  for the motion along $Y$ axis
\begin{eqnarray}
  E_y(0,Y)=0\,,\quad E_x(0,Y) \sim   \frac{\phi_0 }{2\pi c\tau }  \frac{S_y} {Y^2} \,.   
          \label{G(Y)as)}
\end{eqnarray}

Interestingly, the material anisotropy does not enter these results at all. This means that the power-law decay of the electric field should exists also in the isotropic case. In fact, for $\gamma=1$ one has from Eq.\,(\ref{Gr})    \begin{eqnarray}
       G(X,0)=  \frac{i\phi_0S_x}{4\pi^2c\tau}\int  \frac{d^2\bm q \, q_x  e^{i  \bm q\bm X}}{ q^2(1+q^2)} ,\qquad 
   \label{GX}
\end{eqnarray}
which is readily done integrating first over the angle between $\bm q$ and $\bm X$. We obtain:
\begin{eqnarray}
       G(X,0)=  \frac{\phi_0S_x}{2\pi c\tau}\left[ K_1(X) -  \frac{1}{ X}\right] ,\qquad 
   \label{GX1}
\end{eqnarray}
that gives
\begin{eqnarray}
   E_y(X,0) = -\frac{\phi_0 S_x}{2\pi c\tau }  \left[ K_1^\prime(X) + \frac{1}{ X^2}\right] .   
          \label{e44}
\end{eqnarray}
 \begin{figure}[htb]
\includegraphics[width=7cm] {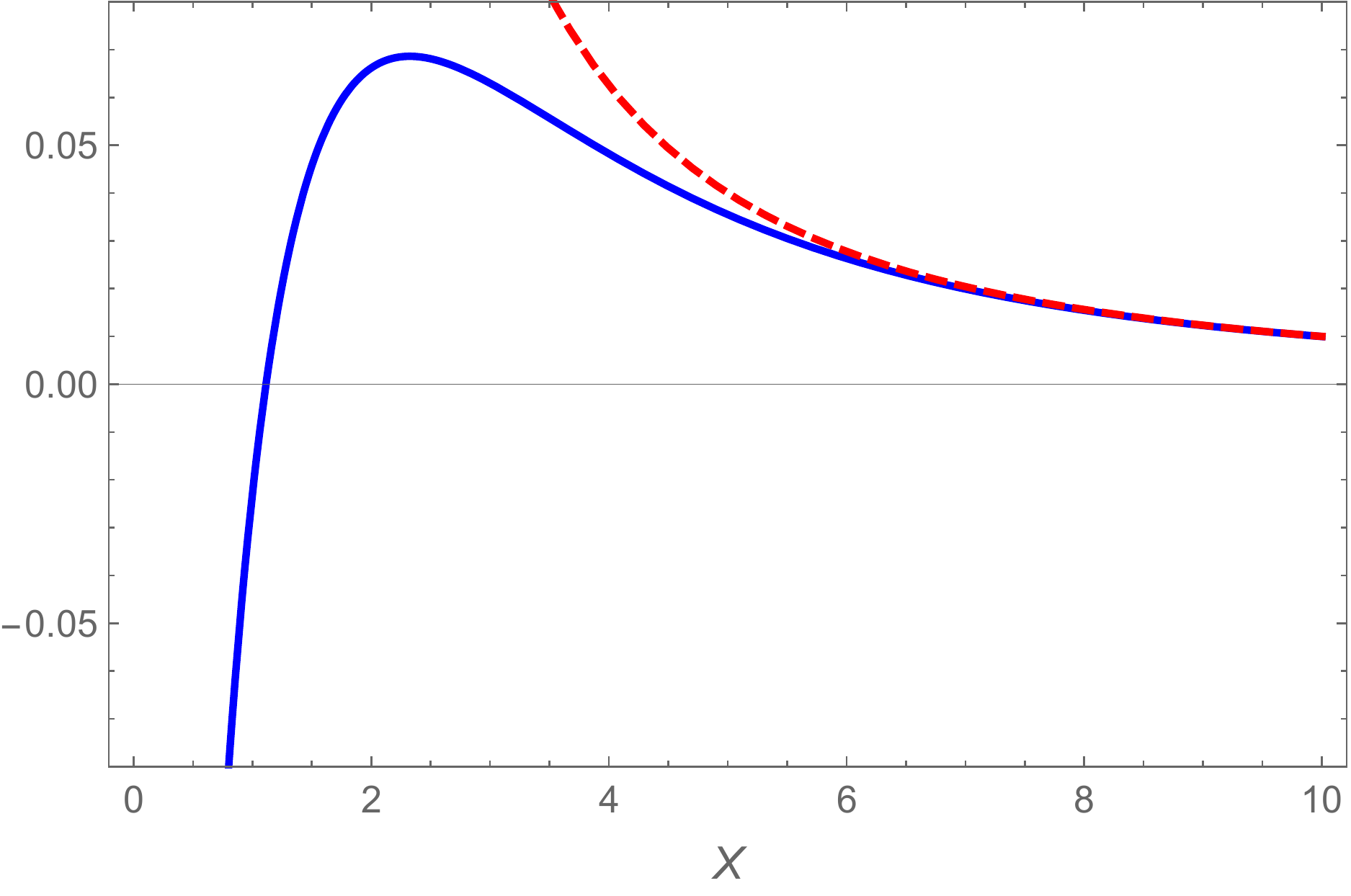}
\caption{ The  solid line is the square brackets in Eq.\,(\ref{e44}) for $E_y(X,0)$ when the vortex moves along    the $X$ axis ($S_y=0$), $\gamma=1$. The dashed line shows  the power-law term $1/X^2$. $X$ are in units of $\lambda$.  }
\label{f10}
\end{figure}

Figure \ref{f10} shows that the field $E_y(X,0)$ changes sign at   $x/\lambda\approx 1$, reaches maximum near 2, and  slowly decays as a power law $\lambda^2/x^2$. This is quite surprising since the electric field   power-law decay means that  no screening of $\bm E$ is involved, in other words, that there is no Meissner-type effect for the field $\bm E$.


\section{Discussion}


We have studied effects of vortex motion within time dependent London theory, which is based on the assumption that  in time dependent phenomena the current in superconductors consists of the persistent and normal components, Eq.\,(\ref{current1}). This approach differs from the common assumption that the vortex magnetic structure moves as a whole, so that in the frame bound to the moving vortex the magnetic field distribution is the same as for a vortex at rest, see e.g. \cite{Dolgov} or  multitude of papers describing the flux flow.


Within the TDL approach the field distribution of the moving vortex differs from that of vortex at rest even in the frame moving  with the vortex. The physical reason for this is simple: the moving magnetic structure $h(x,y)$ induces the electric field and currents of normal excitations, while the latter distort the moving static field distribution $h_0(x,y)$. This is a general feature of systems with singularities (vortices) moving in dissipative media \cite{leo,TDL}. The  equation describing these time-dependent phenomena are diffusion-like, so that solutions are obtained in the  2D Fourier space: we obtain $h_{\bm k}$ and to recover  $h({\bm r})$ one has to evaluate double integrals $\int d^2\bm k...\,$, a heavy numerical procedure. We offer a way to reduce double integrals to a single $\int_0^\infty d\eta...$ which can be evaluated within Wolfram Mathematica package efficiently and fast, that is relevant especially for generating  plots of various 2D distributions.
 
We have investigated the field distribution of moving vortices  away of the vortex core whether the time-dependent London theory   is reliable. 
As in the isotropic case \cite{norio1}, the magnetic field of moving vortices in anisotropic materials is distorted relative to the static case, the magnetic flux is redistributed so that it is depleted in front of the moving vortex and enhanced behind it. The depletion could be  strong enough so that the field $h_z$ changes sign in some parts of the $xy$ plane. This suggests that the interaction of two vortices, one at the origin at some moment and another is at $(x,y)$, being repulsive  at short intervortex distances may turn attractive.

 The physical reason for this change is the induced electric field $\bm E$ and along with it the currents of normal excitations $\sigma \bm E$. This field is obtained by solving quasi-stationary Maxwell equations curl$\bm E=\partial_t\bm h/c$, the condition of quasi-neutrality div$\bm E=0$, coupled with the time-dependent London equation (basically, the same procedure as in deriving time-dependent Ginzburg-Landau equations \cite{Gorkov}). Unlike $\bm h$, the field $\bm E$ cannot be screened in the bulk of the material, so that one may say that there is no ``Meissner effect" for the electric field {\it per se}.

It turns out that in anisotropic case the magnetic field of moving vortex has a power law dependence on distances $r>>\lambda$: $h \propto (\gamma^2-1)v/r^3$ ($\gamma$ is the anisotropy parameter, $v$ is the vortex velocity).
The exponentially decaying part of $h$ is still present, but at large distances it is irrelevant in comparison with the power-law part. 
In isotropic case, the power law gives way to the standard exponential decay. The electric field, however, goes as $1/r^2$ in both cases. 
 
Most of our calculation were done for orthorhombic materials with the in-plane anisotropy parameter $\gamma = 3$ and the vortex along $c$. Such materials in fact exist,   examples are NiBi films \cite{NiBi}, or Ta$_4$Pd$_3$Te$_{16}$ \cite{17}.

\appendix

\widetext 
\section{}

 Consider the  integral
  \begin{eqnarray}
  \int  \frac{d^2\bm k e^{i\bm k \bm r}} {(1+ k ^2 )^2}
=\int_0^\infty \frac{k\, dk}{(1+ k ^2 )^2}\int_0^{2\pi} d\varphi e^{ikr\cos(\alpha-\varphi)}  
 = 2\pi \int_0^\infty \frac{k\, dk}{(1+ k ^2 )^2}J_0(kr) =\pi rK_1(r)\,,
   \label{A1}
\end{eqnarray}
 $\bm k$ and $\bm r$ are at angles $\varphi$ and $\alpha$ relative to $x$. Apply $\partial_x$ to both sides:
 \begin{eqnarray}
 \int  \frac{d^2\bm k k_xe^{i\bm k \bm r}} {(1+ k ^2 )^2}=i\pi xK_0(r)\,,
   \label{A1}
\end{eqnarray}
The evaluation of the first integral in Eq.\,(\ref{corr1}) is now straightforward.

 \section{}
 
 The second contribution in Eq.\,(\ref{corr1}) consists of  parts related to $x$ and $y$ projection of the velocity. With the help of identities
 \begin{eqnarray}
  \frac{1}{f} =\int_0^\infty du\,e^{-fu}\,,\quad  \frac{1}{f^2} =\int_0^\infty du\,u\, e^{-fu},\qquad
   \label{B1} 
\end{eqnarray}
one recasts the $x$-part:

\begin{eqnarray}
I_x=\frac{S_x(1-\gamma^2)}{\gamma}  \int  \frac{d^2\bm q\,q_y^2 q_x  e^{i\bm q \bm R}} 
 {q^2(1+\gamma q_x^2+  q_y^2/\gamma)^2} 
  =  \int_0^\infty d\xi \int_0^\infty du\,ue^{-u}\int d^2\bm q\,q_y^2q_xe^{i\bm q \bm R-(\xi+u\gamma)q_x^2-(\xi+u/\gamma)q_y^2}\,. 
  \label{sx}
\end{eqnarray}
Here we use $\lambda$ as a unit length: $\bm q=\bm k\lambda$, $\bm R=\bm r/\lambda$$, \lambda_2^2=\lambda^2\gamma$,  $\lambda_1^2=\lambda^2/\gamma$, and $\bm S=\bm s /\lambda$. The anisotropy parameter is $\gamma=\lambda_2/\lambda_1$.
We now introduce a new integration variable $\zeta$ via $\xi=\zeta u$: 
 \begin{eqnarray}
I_x  = \frac{S_x(1-\gamma^2)}{\gamma} \int_0^\infty d\zeta \int_0^\infty du\,u^2e^{-u}\int d^2\bm q\,q_y^2q_xe^{i\bm q \bm R-u(\zeta+ \gamma)q_x^2-u(\zeta+ 1/\gamma)q_y^2}\,. 
  \label{B3}
\end{eqnarray}
Integrals over $q_x,q_y $ are evaluated with the help of the known Fourier transform of a Gaussian: 
\begin{eqnarray}
  \int_{-\infty}^\infty   dq_x\, e^{i q_x x-a q_x^2  }  =\sqrt{\frac{\pi}{a}}e^{-x^2/4a}.\qquad
\label{int} 
\end{eqnarray}
 
Integration over $u$ can be done utilizing relations
\begin{eqnarray}
  \int_0^\infty  \frac{du}{u}  \, \exp\left(-u-\frac{w^2}{4u}  \right)  =2\,K_0(w)\,,\qquad  
   \int_0^\infty  \frac{du}{u^2}  \, \exp\left(-u-\frac{w^2}{4u}  \right)  =\frac{4}{w}\,K_1(w)\,.
  \label{B5} 
\end{eqnarray}

We obtain after straightforward algebra:
 \begin{eqnarray}
I_x =  \frac{i\pi (1-\gamma^2)}{2\gamma} S_x X
 \int_0^\infty \frac{d\zeta }{(\zeta+ \gamma)^{3/2}(\zeta+ 1/\gamma )^{3/2}} 
  \left[ K_0(R_\zeta)-\frac{Y^2}{(\zeta+1/\gamma)R_\zeta}K_1(R_\zeta) \right]\,, \quad  
 R_\zeta^2=\frac{X^2}{\zeta+\gamma}+\frac{Y^2}{\zeta+1/\gamma}\,.\qquad
  \label{B6}
\end{eqnarray}
In a similar fashion one obtains the part proportional to $S_y$ and Eq.\,(\ref{dh2}). 

\section{Electric field and normal currents}

We  evaluate   here the stream function of Eq.\,(\ref{Gr}):
\begin{equation}
G=\frac{i\phi_0c}{4\pi^2 c\tau } \,{\hat G}\,,\quad   {\hat G} =\int \frac{d^2\bm q (\bm q\cdot\bm S)}{ q^2C} e^{i\bm q \bm R}\,.   
    \label{C1}
\end{equation}
 The following manipulation is similar to that outlined in Appendicx B for $h(X,Y)$:
\begin{eqnarray}
{\hat G} =   \int d^2\bm q(\bm q\cdot\bm S) e^{i\bm q\bm R}  \int_0^\infty du   e^{-uq^2}  \int_0^\infty d\xi e^{-\xi C}
=\int_0^\infty du \int_0^\infty d\xi  e^{-\xi} \int d^2\bm q (\bm q\cdot\bm S) e^{i\bm q\bm R-uq^2-\xi(q_x^2\gamma+q_y^2/\gamma)}. 
\label{C2}
\end{eqnarray}
Further, we write the last integral 
$\int d^2\bm q ...  =S_xI_x+S_yI_y$ with 
\begin{eqnarray}
I_x =   \int_{-\infty}^\infty dq_x q_x e^{i q_x X  - q_x^2(u+\xi\gamma)} \int_{-\infty}^\infty dq_y   e^{i q_yY  - q_y^2(u+\xi/\gamma)}   
\label{C3}
\end{eqnarray}
and $I_y$ which is obtained from $I_x$ by replacing $x\leftrightarrow y$ and $\gamma \leftrightarrow 1/\gamma$. The integral over $q_x$ and $q_y$ are related to the known Fourier transform of a Gaussian and we obtain:
\begin{eqnarray}
I_x =   \frac{i\pi X}{2(u+\xi\gamma)^{3/2}(u+\xi/\gamma)^{1/2}}  \exp\left(-\frac{X^2}{4(u+\xi\gamma)}-\frac{Y^2}{4(u+\xi/\gamma)}\right)    
\label{C4}
\end{eqnarray}
and the part $\hat G$ proportional to $S_x$ takes the form
\begin{eqnarray}
 {\hat G}_x=   \frac{i\pi XS_x}{2}\int_0^\infty du     \int_0^\infty \frac{d\xi \,e^{-\xi}  }{(u+\xi\gamma)^{3/2}(u+\xi/\gamma)^{1/2}} \exp\left(-\frac{X^2}{4(u+\xi\gamma)}-\frac{Y^2}{4(u+\xi/\gamma)}\right) ,\quad   
\label{C5}
\end{eqnarray}
 
To integrate over $u$ we can use  Eq.\,(\ref{B5}).    
To this end we introduce a new integration variable $\eta$ instead of $\xi$ via $\xi=u\eta$. Then the integral over $\xi$ becomes
 \begin{eqnarray}
\frac{1}{u}   \int_0^\infty d\eta \frac{  \, e^{-\eta u}}{ (1+\eta\gamma)^{3/2}(1+\eta/\gamma)^{1/2}}   \exp\left( -\frac{{\cal R}_\eta^2}{4u }\right)\,,\quad {\cal R}_\eta^2=\frac{X^2}{1+\eta\gamma }+\frac{Y^2}{ 1+\eta/\gamma }.
   \label{C7}
\end{eqnarray}
 Now, the integration over $u$ is done with the help of Eq.\,(\ref{B5}) and we obtain:
\begin{eqnarray}
 {\hat G}_x =i  \pi   S_xX  
   \int_0^\infty  \frac{d\eta  }{ (1+\eta \gamma)^{1/2}(1+\eta /\gamma)^{3/2}}  \,K_0(R_\eta \sqrt{ \eta }) .
      \label{C8}
\end{eqnarray}
The part $G_y$ follows immediately after the replacements $x \leftrightarrow y$ and $(1+\eta\gamma) \leftrightarrow (1+\eta/\gamma)$: 
\begin{eqnarray}
 {\hat G}_y =i  \pi   S_yY  
   \int_0^\infty  \frac{d\eta  }{ (1+\eta \gamma)^{3/2}(1+\eta /\gamma)^{1/2}}  \,K_0({\cal R}_\eta \sqrt{ \eta }) .
      \label{C8}
\end{eqnarray}


\begin{thebibliography}{99}


\bibitem{Grishin} A. M. Grishin, A. Yu. Martynovich, S. V. Yampolsky, Zh. Eksp. Teor. Fiz. {\bf 96}, 1930 (1990).
 
\bibitem{Buzdin} A. I. Buzdin and A. Yu. Simonov, Pis'ma Zh. Eksp. Teor. Fiz.
51, 168 (1990) [JETP Lett. 51, 191 (1990)].

\bibitem{NakThiem} V. G. Kogan,  N. Nakagawa, S. L. Thiemann, \prb  {\bf 42},  2631 (1990).

\bibitem{Bolle}C. A. Bolle, P. L. Gammel, D. G. Grier, C. A. Murray,  D. J. Bishop, D. B. Mitzi and A. Kapitulnik, \prl {\bf 66}, 112 (1991).
 
\bibitem{forces} V. G. Kogan,   \prl  {\bf 64},  2192 (1990).

 \bibitem{leo}L. Radzihovsky,   Phys. Rev. Lett. {\bf 115}, 247801 (2015). DOI: 10.1103/PhysRevB.97.094510. DOI: 10.1103/PhysRevLett.115.247801

\bibitem{TDL}  V. G. Kogan, \prb {\bf 97}, 094510 (2018). 


\bibitem{Andreev} M. Smith, A. V. Andreev, and B. Z. Spivak,   \prb {\bf 101}, 134508 (2020). 

\bibitem{Maeda} R. Ogawa,  F. Nabeshima,  T. Nishizaki,  and A. Maeda, \prb {\bf 104}, L020503 (2021); arXiv:2105.15118.

\bibitem{K81} V. G. Kogan,   \prb  {\bf 24}, 1572 (1981). 

\bibitem{anisTDL}V. G. Kogan and R. Prozorov, \prb {\bf 102}, 184514 (2020).

\bibitem{LL} L. D. Landau, E. M. Lifshitz, and L. P. Pitaevskii, {\it Elecrtodynamics
of Continuous Media}, 2nd ed. (Elsevier, Amsterdam,1984).

\bibitem{Gorkov}  L. P. Gor'kov and N. B. Kopnin, Usp. Fiz. Nauk, {\bf 116}, 413 (1975); Sov. Phys.-Usp., {\bf 18}, 496 (1976).


 \bibitem{Eli}L. Embon, Y. Anahory, \v{Z}.L. Jeli\'{c}, E.O.Lachman, Y. Myasoedov,
 M. E. Huber, G. P. Mikitik, A. V. Silhanek, M. V. Milosevi\'{c}, A.
 Gurevich, and E. Zeldov,   Nat. Commun. 8, 85 (2017).

 \bibitem{Denis}O. V. Dobrovolskiy,  D. Yu. Vodolazov,  F. Porrati,  R. Sachser, 
 V. M. Bevz,  M. Yu. Mikhailov,  A. V. Chumak,  and M. Huth,   Nature Communications {\bf 11}, 3291.


\bibitem{norio1}  V. G. Kogan and N. Nakagawa, Condense Matter,  {\bf 4}, 6 (2021).  https://doi.org/10.3390/condmat6010004.

\bibitem{Norio2}  V. G. Kogan and N. Nakagawa, \prb {\bf 103}, 134511 (2021). 

\bibitem{Dolgov} O. V. Dolgov and N. Schopohl, \prb {\bf 61}, 12389 (2000).



\bibitem{NiBi}Wen-Lin Wang, Yi-Min Zhang,  Yan-Feng Lv,  Hao Ding,  Lili Wang,  Wei Li,  KeHe,  Can-LiSong,  Xu-CunMa,  and Qi-KunXue, arXiv:1804.09890 (2018). 


\bibitem{17}Y. Fujimori, S. I. Kan, B. Shinozaki, and T. Kawaguti, J.
Phys. Soc. Jpn. {\bf 69}, 3017 (2000).
 
\end{thebibliography}
\end{document}